\documentclass[10pt,twocolumn,english,pre]{revtex4}

\usepackage{amsmath}    
\usepackage{graphicx}   

\usepackage[T1]{fontenc}
\usepackage[latin1]{inputenc}
\usepackage{graphicx,epsfig}
\usepackage{html}
\usepackage{natbib}

\begin{document}
\def\deg{^\circ}
\def\AA{{\rm A}}
\def\half{ \small {1 \over 2}}
\bibliographystyle{./simon}

\title{Biologic: Gene circuits and feedback in
an 
introductory physics
sequence 
for biology and premedical students
}

\author{S. B. Cahn}
\affiliation{Department of Physics, Yale University, New Haven, Connecticut 06511}
\author{S. G. J. Mochrie}
\affiliation{Department of Physics, Yale University, New Haven, Connecticut 06511}
\affiliation{Department of Applied Physics, Yale University, New Haven, Connecticut 06511}

\date{\today}

\pacs{pacs}

\begin{abstract}
Two synthetic gene circuits
-- the genetic toggle switch and the repressilator --
are
discussed in the context
of an educational module on gene circuits and feedback
that constitutes the final topic of a year-long introductory physics sequence,
aimed at biology and premedical undergraduate students. 
The genetic toggle switch consists of two genes, each of whose protein product
represses the other's expression, while
the repressilator consists of three genes,
each of whose protein product represses the next gene's expression.
Analytic, numerical, and electronic treatments of the genetic toggle switch
shows that this gene circuit realizes bistability.
A simplified treatment of the repressilator reveals that this circuit can realize sustained oscillations.
In both cases,  a ``phase diagram'' is obtained, that specifies the region of parameter space
in which bistability or oscillatory behavior, respectively, occurs.
\end{abstract}

\maketitle

\section{Background and Introduction}
Two recent reports -- the NRC's
``BIO2010: Transforming Undergraduate Education
for Future Research Biologists'',
\cite{BIO2010}
and the AAMC/HHMI's
``Scientific Foundations for Future Physicians'' \cite{AAMC} --
have highlighted the increasing importance of quantitative skills
for students who are planning biomedical careers. 
Since the 2010-2011 academic year, the Yale physics
department has offered a new introductory physics sequence -- PHYS 170/171 --
aimed at  biology and premedical students,
that seeks to implement a number of the recommendations of these
reports.
The PHYS 170/171 enrollment was about 100 in the 2010-2011 academic year,
but increased to nearly 150 in 2012-2013.
The majority of the PHYS 170/171 class (70\%)  are biology majors,
 and 80\%  identify themselves as premedical students.
There are roughly equal numbers of sophomores and juniors, with
significantly fewer seniors, and two or three freshmen.
They are 70\% female.  50\% self-identify as white,
50\% do not.
Most come to PHYS 170/171
possessing considerable biological sophistication,
because of prior biology and chemistry classes at Yale.
Almost all have previously taken a first course in calculus.

In considering a new introductory physics syllabus for 
biology and premedical students, there is a tension
between what topics seem likely to be interesting and engaging
versus what has traditionally been
covered in introductory physics courses.
Our starting point for resolving the PHYS 170/171  syllabus
is the observation that the majority
of these students will not take another physics course after PHYS 170/171.
Therefore, we reasoned,  there is no rationale
to prepare students
for more advanced classes
or the physics major.
Instead, our selection of topics and our approach is informed by the desire to
tackle interesting topics, that demonstrate that
physics has much to contribute to the life sciences and medicine,
and that reflect that
biological physics is now a major sub-field of physics, well-represented in
physics departments across the country.
The principle that we should endeavor to align
physics teaching
with how we practice physics also resonates with us.\cite{Handelsman2007}
Thus, we have been led to include modules on chemical rate equations, probability,
Brownian motion and diffusion,
laminar fluid flow, statistical mechanics and Brownian ratchets,\cite{Mochrie2011}
electromagnetic waves,  and quantum mechanics -- all topics which, between the two of us,
we engage with in our own research.

In this paper, we present our PHYS 170/171 module on gene circuits
in the hope that it will prove useful to
others also thinking about new physics curricula for biology and premedical undergraduates.
We decided to include a
gene circuits module  --
humorously called ``Biologic'' -- because of the importance of the concept of feedback to clinicians,
and to provide an introduction to biological control, decision making, and time-keeping,
which constitute the subject matter of  ``Systems Biology'', which has emerged as
a major subfield of biology over the last decade,
and to which physicists and engineers have made key contributions.
Our educational goals are to introduce and explore the concept of feedback,
to show that feedback can lead to switches and oscillators, both in electronic circuits and in gene circuits, 
and to alert students to the place of quantitative approaches in Systems  Biology.
To this end, we present simplified treatments of two
{\em de novo} designed gene circuits, namely
 the ``genetic toggle switch'' \cite{GTS} and the ``repressilator'',\cite{REPRESS} each of
which has been realized experimentally in {\em E. coli}.
Nature relies on multiple,  interconnected gene circuits,
  that are considerably more refined than these
 Frankensteinian examples.
Nevertheless,  analogous natural gene networks are ubiquitous.
For example,  in the case of the genetic toggle switch,
there is an analogy with the well-studied lytic-lysogenic switch in the bacteriophage lambda life cycle
\cite{Ptashne}, and a genetic oscillator governs the development of vertebrate
segmentation that eventually leads to vertebrae.\cite{HolleyReview}
A version of the repressilator has recently been recognized in the gene circuit of {\em Bacillus subtilis},
a common soil bacterium.\cite{Schultz2009}
 
''Biologic'' is the final module in the year-long PHYS 170/171 introductory physics sequence.
By this time, students' mathematical skills have been practiced by nearly two semesters of physics.
From earlier modules, they are familiar with coupled harmonic oscillators, eigenvalues and eigenvectors,
and rate equations for chemical reactions,
and they have had considerable experience with Wolfram Alpha and Wolfram Demonstrations.
Throughout the year, we emphasize using Wolfram Alpha to facilitate mathematical
manipulations,
including the solution of systems of algebraic equations,
the evaluation of derivatives and integrals, and
the numerical solution of differential equations,
which we believe empowers the students.
Because computational approaches constitute an essential aspect of how
research is now carried out, both in the physical  and  life sciences,
we also include
a number of
simulations and visualizations,\cite{Dori2003,Chabay2008}
implemented as Wolfram Demonstrations,\cite{DEMOSITE}
which run in students' favorite web browsers.
%
Students' positive responses to Mathematica Demonstrations
assuaged initial doubts amongst faculty colleagues, concerning
the students' ability and willingness to use such software.

In class, we segue from the previous module -- electromagnetic waves -- to genetic circuits
by invoking what is arguably the twentieth century's greatest invention, namely the transistor.
We point out that the proliferation of transistors, which rely on Maxwell's equations for their function,
continues to transform the way we live, and assert that transistors are electronic switches.
We further point out that, just as electronic switches and circuits implement
electronic ``decisions'' depending on certain inputs, analogously biology uses
biological switches and circuits to implement biological decisions.

To emphasize the intellectual connection between gene circuits and electronic circuits,
in the associated laboratory course, PHYS 165/166, we implemented
electronic versions of the genetic toggle switch and the repressilator.
The electronic toggle switch was built using two of the logical inverters of a
7404 Hex Inverter.
The electronic repressilator was built using three NAND gates of 
a 7400 Quad NAND chip, together with appropriate resistors and capacitors to select the oscillation
period.
The PHYS 165/166 laboratory handout
is included in the Supplementary Information.
To provide further opportunities for exploration, we exploit that Mathematica can numerically solve the
relevant equations both for the genetic toggle switch and the repressilator.
These solutions are presented in Wolfram Demonstrations.\cite{gtsdemo,repressdemo}

\section{Phage Lambda: Lysogeny or lysis, that is the question}
From their biology classes, many PHYS 170/171 students are familiar with
 the life cycle of bateriophage lambda, which realizes one of the most studied and best understood biological
 switches, between the so-called lytic and lysogenic states.\cite{Ptashne}
 For infected bacteria in the lysogenic state, the bacteriophage's genetic material
 is incorporated in the bacterial chromosome and
 is replicated along with the host's genetic material at cell division,
 but phage capsid proteins, {\em etc.} are not expressed.
However, in the lytic state, the proteins required to form new phage are expressed,
many copies of the phage assemble, and the
the host is caused to disintegrate (lyse),
releasing many new bacteriophage particles, free now to infect a new host.  The write-up for the laboratory module describes that in the lytic state, the phage
 make lots of copies of themselves and their {\em spacesuits};  then they blow up their host bacterium and disperse,
 presumably to find the next victim.
These two different possible outcomes are visualized in
the YouTube movie,
\href{http://www.youtube.com/watch?v=sLkZ9FPHJGM}{\tt http://www.youtube.com/v/sLkZ9FPHJGM},
 in which phage infection is signaled by green fluorescence.\cite{Pierre2008}
In the top right of the field of view, early on in the movie, a replicating {\em E. coli}
becomes infected and undergoes lysis, providing an example of the behavior in the lytic state.
By contrast, in the lower left of the field of view, another {\em E. coli} becomes infected, but, instead of undergoing lysis,
this bacterium undergoes two rounds of cell division before the movie ends, resulting
in four infected, fluorescent, daughter {\em E. coli} in the lysogenic state.

The existence of two possible outcomes in this system, {\em i.e.} bistability,
depends on
 the interaction of two genes, {\em cI} and {\em cro}
 and their protein products, cI and cro, each of which represses the other.
In the lysogenic state,  the concentration of cI is high,  
expression of cro is repressed, and expression of cI continues at a high level,
which continues to repress cro, and so on. 
Alternatively, in the lytic state, cro is expressed, which represses cI, which therefore is unable to repress cro,
therefore cro remains high
and the lytic state persists, or would persist, except that this state initiates a pathway to host cell lysis.
We note that here feedback is exemplified by the protein product of a certain gene
then going on to affect in some way its own expression. 
In addition to repressing cI, cro is also an activator for cro,
realizing a second positive feedback loop, which
represents the ``suspenders'' in a ``belt-and-suspenders'' approach
to maintaining the lysogenic state.
 

\section{Genetic toggle switch}
\begin{figure}[t!]
\vspace{-0.1in}
\begin{center}
{\includegraphics[width=0.35\textwidth,keepaspectratio=true]{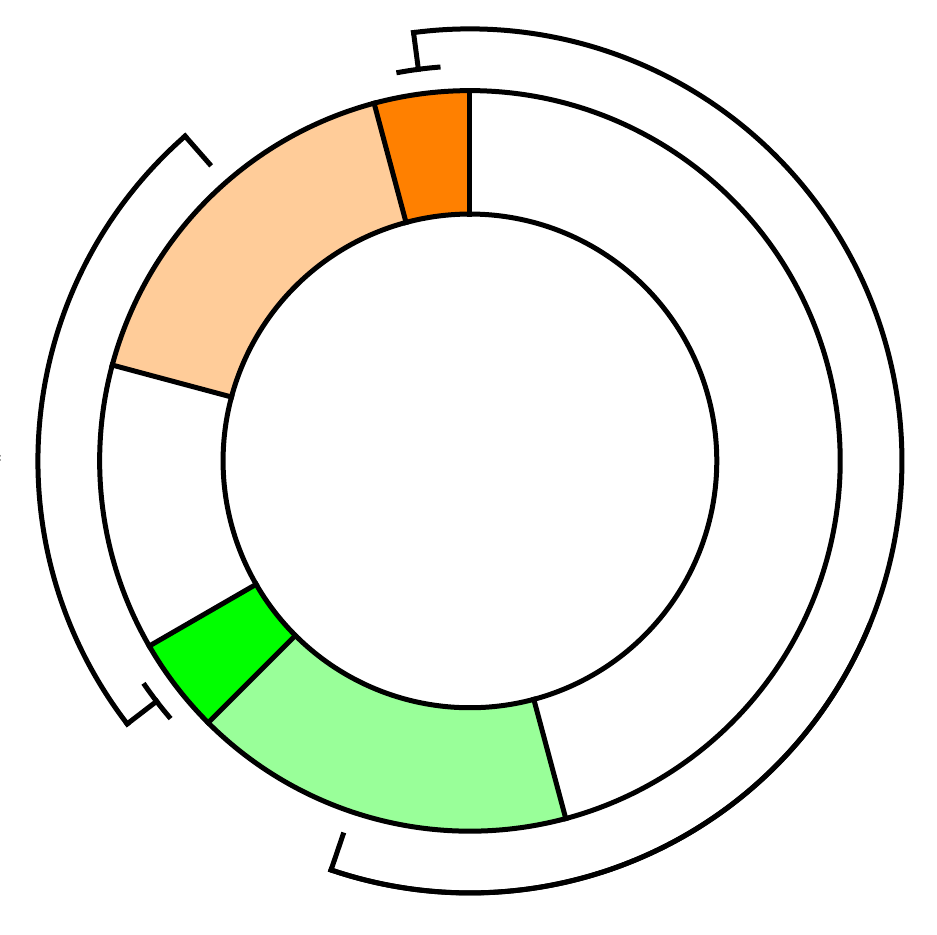}}
\end{center}
\vspace{-0.2in}
\caption{
Schematic of the genetic toggle switch plasmid. The solid orange region is the promoter for
repressor 1.
The solid green region is the promoter for repressor 2.
The light orange region codes for repressor 1, which, as indicated,
represses expression of repressor 2. The light green region
codes for repressor 2, which, as indicated, represses promoter 1.
}
\label{GTSplasmid}
\vspace{-0.1in}
\end{figure}

The design of the genetic toggle switch is shown in Fig. \ref{GTSplasmid}.
The genetic toggle switch is an even simpler bistable gene circuit than the lambda switch,
consisting of two genes each of which encodes for a protein that represses the other's gene expression.
Because of the even number of components in the circuit, there is a net positive feedback,
which can give rise to bistable behavior for certain parameter values. 
The genetic toggle switch was described and implemented in {\em E. coli} in Ref. \onlinecite{GTS}.

\subsection{Electronic realization of a toggle switch}
\begin{figure}[t!]
\begin{center}
\includegraphics[width=3in]{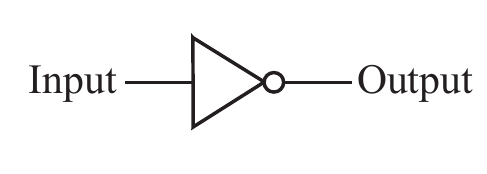}
\includegraphics[width=3in]{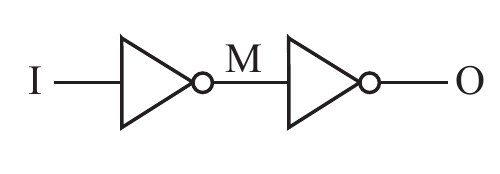}
\includegraphics[width=3in]{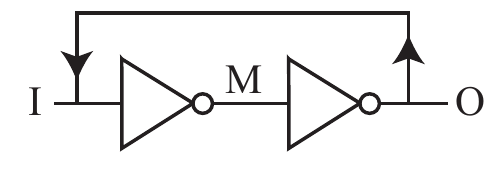}
\end{center}
\caption{Top: Depiction of a logical inverter. The circle represents inversion.
Center:  Two cascaded investors.
Bottom: Two cascaded inverters with feedback.
}
\label{inverter}
\end{figure}

The concept of bistability and the role of feedback is emphasized
in the laboratory component of the course by first introducing and explaining
the operation of a logical inverter, shown at the top of Fig.~\ref{inverter}.
We explain that the logical inverter realizes the operation that changes {\tt True} to {\tt False}, yes to no,
1 to 0, high to low and {\tt False} to {\tt True}, no to yes, 0 to 1, low to high.
We then consider two inverters in series (center of Fig. \ref{inverter}).
In this case,  if the Input (I) is {\tt True},
the intermediate result (M) is {\tt False},
and the Output (O) is {\tt True}. 
Likewise, if I is {\tt False}, M is {\tt True}, and O is {\tt False}.
Finally, we ask students to consider what happens when
we introduce feedback, that is, where we take the Output
and route it back to the Input, as shown at the bottom of  Fig.~\ref{inverter}.
Here, if I is {\tt False}, M is {\tt True}, O is {\tt False}, which feeds back to I, which was already {\tt False}.
Thus, we see that our ``circular'' logic is self-consistent in this case.
Even if we were to remove the input I, the logic speedway would remain stable.
But what if we
force M to be {\tt False}, which makes O {\tt True}, which renders I {\tt True} and M {\tt False}.
Our logic is consistent in this case too.
Both conditions are stable, so that two logical inverters with feedback realize a bistable situation.
The Output can be set to either value, and it will stay in that state indefinitely,
{\em i.e.}, it is a Toggle Switch.
We point out that such a structure can also be considered a memory element,
with a value at the Output of {\tt True} or {\tt False}, 0 or 1, a binary digit or bit.

\begin{figure}[t!]
\begin{center}
\includegraphics[width=3in]{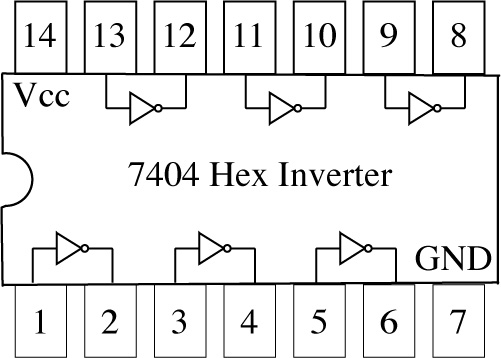}
\vskip 0.5 in
\includegraphics[width=3in]{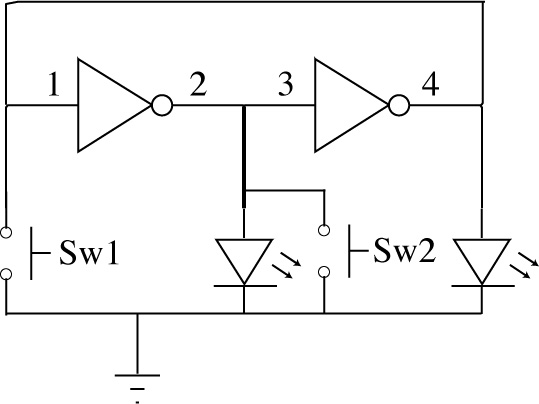}

\end{center}
\caption{Top: 7404 Hex Inverter Chip.
Bottom: Realization of electronic toggle switch with 7404}
\label{7404}
\end{figure}

To actually implement an electronic toggle switch in the laboratory module, we use two of the
logical inverters of a 7404 hex inverter chip (Fig. \ref{7404}). Students are provided with
the chip already appropriately wired up to use the first two logical inverters only:
The chip is mounted on a breadboard,  is powered with $V_{CC}=5$~V at pin 14 and is grounded at pin 7.
To construct the feedback circuit, pin 2 is connected to pin 3 and pin 4 is connected to pin 1.
In addition,  the chip is connected to two ``Morse Code'' switches, which can bring pin 1 to ground (False)
or pin 3 to ground. (The molecular biological analogues of these switches are so-called inducers.)
The outputs, pins 2 and 4, are also connected to LED indicators, 
which are illuminated when the voltage at pin 2 or 4 is high, to provide a readout of the state of the circuit.
Students are asked to  
sketch the circuit in their laboratory notebooks. They are asked
what happens when you push one switch, and then push it again,
what happens when they push the other switch, and whether this circuit
follows the bistable ``logic'' described above.
For most of the students, this exercise represents the first contact they have had with Boolean logic, and with digital electronics..... beyond their role as consumers, of course.

\subsection{Chemical rate equations for the genetic toggle switch}
An important theme throughout PHYS 170/171 is that physics is concerned with providing
mathematical descriptions of the natural world.
Therefore, in the lecture portion of the class,  building on earlier modules
on chemical rate equations, and on coupled harmonic oscillators,
we analyze the genetic toggle switch as follows:
First, we  write down chemical rate equations for the concentrations of the two repressors.
Next, we look for and find steady-state solutions to these equations.
Then, we determine which of the steady-state solutions are {\em unstable} and which are {\em stable}. 
Finally, we interpret the stable solutions that we find, and examine how these solutions depend on the model's parameters.

The chemical rate equations that we write to describe the concentrations of repressor 1 ($c_1$) and repressor 2 ($c_2$)  are:
\begin{equation}
\frac{{\rm d\,}c_1}{{\rm d\,}t} = -K_1 c_1 + \gamma_1 [1-P_1(c_2)].
\label{RATE1}
\end{equation}
and
\begin{equation}
\frac{{\rm d\,}c_2}{{\rm d\,}t} = -K_2 c_2 + \gamma_2 [1-P_2(c_1)],
\label{RATE2}
\end{equation}
where
$c_1$ is the concentration of repressor 1,
$c_2$ is the concentration of repressor 2,
 $K_1$ is the degradation rate of repressor 1,
$K_2$ is the degradation rate of repressor 2,
 $\gamma_1$ is the production rate of repressor 1 in the absence of any repression by repressor 2,
 $\gamma_2$ is the production rate of repressor 2 in the absence of any repression by repressor 1,
$P_1(c_2)$ is the probability that repressor 2 binds promoter 1,
and
 $P_2(c_1)$ is the probability that repressor 1 binds promoter 2, thus repressing
expression of repressor 2.
Only that fraction of promoter 2 sites that are not occupied by repressor 1 can bind RNA polymerase and give rise to
repressor 2 expression. Therefore, the production rate of repressor 2
in the presence of a concentration, $c_1$, of repressor 1 is actually $\gamma_2[1-P_2(c_1)]$.
Similarly, the production
rate of repressor 1
in the presence of a concentration, $c_2$, of repressor 2 is actually $\gamma_1[1-P_1(c_2)]$.
Of course, a number of biological processes -- transcription, RNA processing, RNA export, translation {\em etc.} --  are subsumed into these equations and their parameters, but they do
articulate the quote that
``Everything should be made as simple as possible, but not simpler'',
attributed to Albert Einstein.

For analytic simplicity, we will take $K_1=K_2$, $\gamma_1=\gamma_2$ and $P_1=P_2=P$ with
\begin{equation}
P(c_1)=\frac{(c_1/c_0)^n}{1+(c_1/c_0)^n}.
\label{HILL}
\end{equation}
EQ. \ref{HILL} is a so-called Hill function and is an approximate representation of
the concentration dependence of the binding probability and where $c_0$ is the repressor concentration for one-half occupancy of the promoter sites.
A value of $n$ greater than 1 generally represents  cooperativity.
Roughly speaking, the larger the
value of $n$, the greater is the degree of cooperativity.
An example of cooperativity, well-known to the students,
occurs for oxygen binding by hemoglobin, which permits oxygen uptake in the lungs where the
concentration of oxygen in blood is high and oxygen release where the concentration of oxygen is low and oxygen is needed in the body.
In that case, a Hill function (EQ. \ref{HILL})
with  $n=4$ is often used to describe the probability that hemoglobin binds four oxygen molecules. 
Importantly, in order to
realize bistable behavior in the genetic toggle switch some cooperativity is required,
{\em i.e.} $n$ must be greater than 1.

With the simplifications described above, EQ. \ref{RATE1} and EQ. \ref{RATE2} become
\begin{equation}
\frac{1}{K} \frac{{\rm d\,}x}{{\rm d\,}t} =  - x +  \frac{a}{1+y^n} 
\label{RATE1B}
\end{equation}
and
\begin{equation}
\frac{1}{K} \frac{{\rm d\,}y}{{\rm d\,}t} =  -y + \frac{a}{1+x^n} ,
\label{RATE2B}
\end{equation}
where $x = c_1/c_0$, $y = c_2/c_0$, and $a = {\gamma}/{(K c_0)}$.

\subsection{Mathematica solutions for the genetic toggle switch}
Mathematica can numerically solve these equations
(EQ. \ref{RATE1B} and EQ. \ref{RATE2B}) versus time for specified parameters and
 initial
conditions.
The solution is presented to the class as a Mathematica Demonstration.\cite{gtsdemo}
\begin{figure}[h!]
\begin{center}
\includegraphics[scale=0.65]{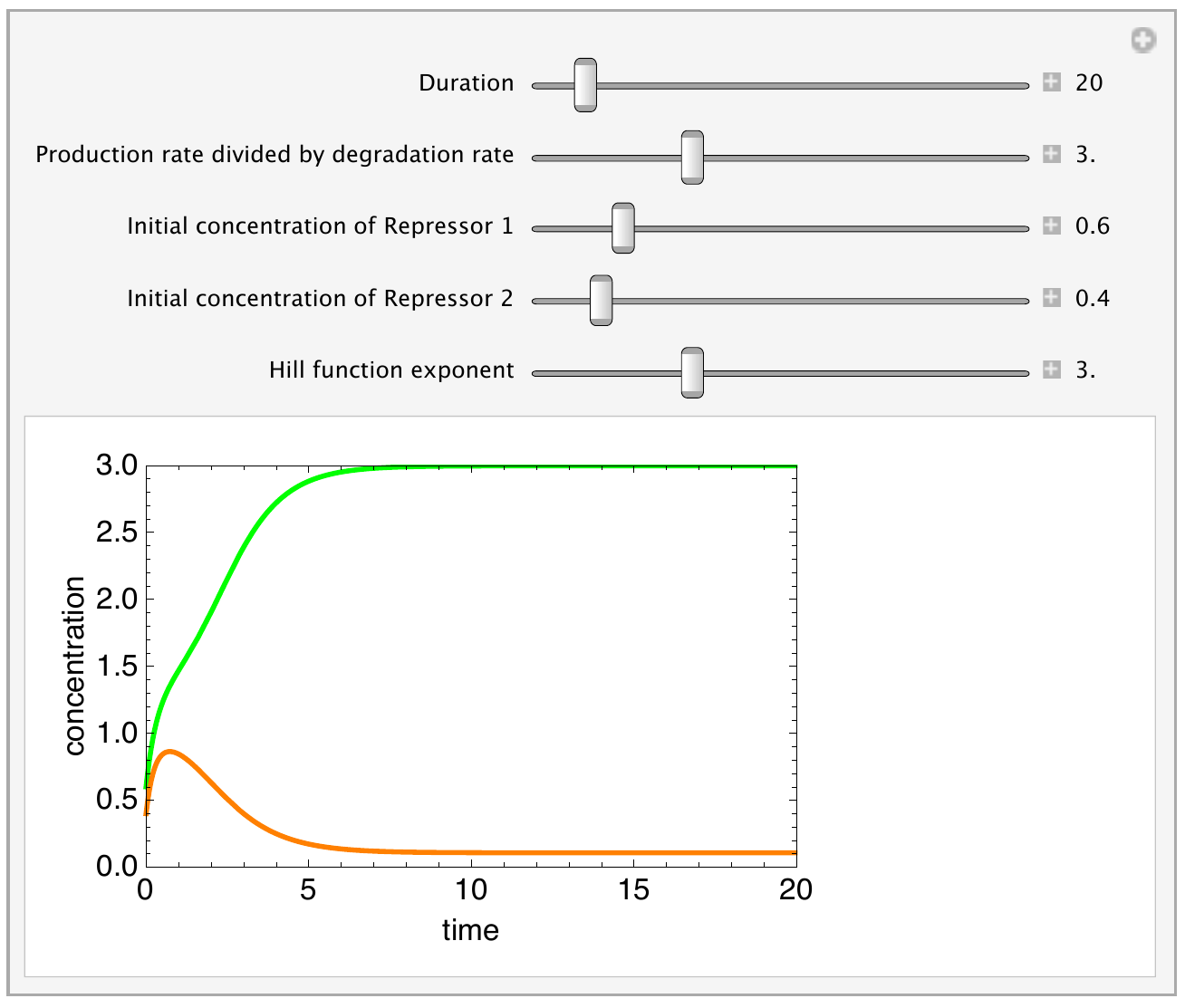}
\end{center}
\caption{Output of the Mathematica Demonstration
\cite{gtsdemo} that solves
EQ. \ref{RATE1B} and EQ. \ref{RATE2B} for the
genetic toggle switch. In this case, $\gamma/(K c_0)=3.0$, 
the initial concentrations are $x(0)=0.6$ and $y(0)=0.4$, and
the time axis is actually $Kt$.
}
\label{GeneticToggleDemo1}
\end{figure}
By varying the sliders in this demonstration, it is possible to vary the parameters of the model and the initial
conditions. This exercise reveals that for some parameters there is a steady-state bistable solution, where  $x$ (green) is large and
$y$ (orange) is small, as shown in Fig. \ref{GeneticToggleDemo1}, or {\em vice versa}, and that it is possible to switch between these two solutions by changing only the
initial conditions.
For other parameters, there is a steady-state solution with $x=y$, irrespective of the initial conditions.
\subsection{Steady-state solutions for the genetic toggle switch}
\begin{figure}[t!]
\begin{center}
\includegraphics[scale=0.96]{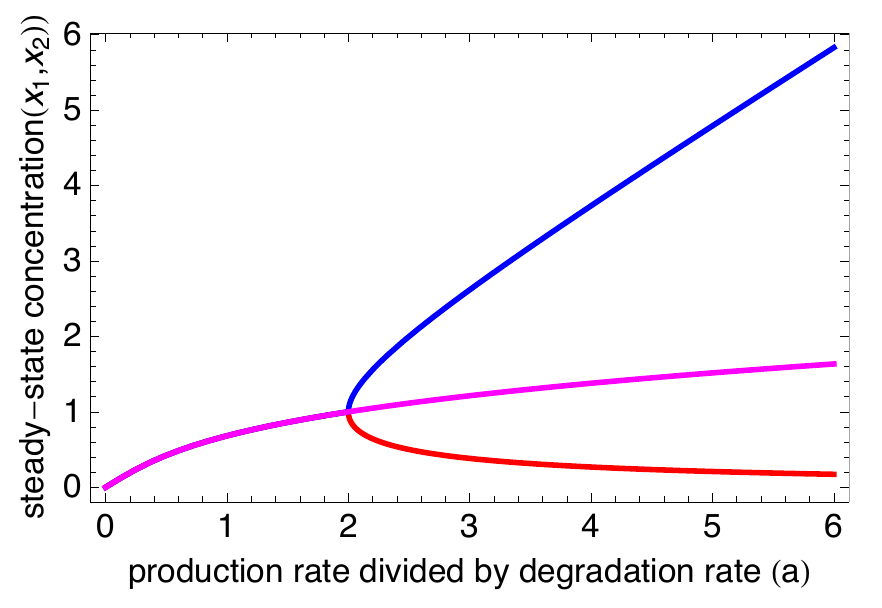}
\end{center}
\caption{The three real solutions of EQ.  \ref{SS1A} and EQ. \ref{SS1B}
plotted versus the (normalized)
repressor production rate divided by the repressor
degradation rate [$a = {\gamma}/{(K c_0)}$].
The two bistable solutions are plotted in blue and red. The other solution is
plotted in magenta.
}
\label{ROOTS}
\end{figure}

For any parameters and initial conditions,  it is evident from the Mathematica
demonstration that the concentrations,
$x$ and $y$,  approach constant values at long times, $x^*$ and $y^*$, respectively.
These values  are the steady-state solutions of EQ. \ref{RATE1B} and EQ. \ref{RATE2B} , defined via
\begin{equation}
 - x^* +  \frac{a}{1+(y^*)^n} =0
\label{SS1A}
\end{equation}
and
\begin{equation}
 -y^* + \frac{a}{1+(x^*)^n} =0.
\label{SS1B}
\end{equation}
In order to proceed analytically,\cite{PBC} we specialize to the case $n=2$, in which case
%
WolframAlpha can solve EQ. \ref{SS1A} and EQ. \ref{SS1B}.
First, we consider non-bistable solutions for which $x=y$.
In this case, both
EQ. \ref{SS1A} and EQ. \ref{SS1B} for $n=2$ become equivalent to:
\begin{equation}
 -x^* + \frac{a}{1+(x^*)^2} =0.
\label{SS1C}
\end{equation}
To solve EQ. \ref{SS1C}, we
navigate to the Wolfram Alpha website.
\href{http://www.wolframalpha.com}{\tt http://www.wolframalpha.com},\cite{WolframAlpha} and enter:
\begin{verbatim}
solve -x+a/(1+x^2)=0 for x
\end{verbatim}
to find:
\begin{equation}
x^*=\frac{(\sqrt{3}\sqrt{27a^2+4}+9a)^{\frac{1}{3}}}{2^\frac{1}{3}3^\frac{2}{3}}
-
\frac{(2/3)^\frac{1}{3}}{(\sqrt{3}\sqrt{27a^2+4}+9a)^{\frac{1}{3}}}.
\label{solution1}
\end{equation}
To solve EQ. \ref{SS1A} and EQ. \ref{SS1B}
in the case that  $x \neq y$, we  enter
\begin{widetext}
\begin{verbatim}
solve -x+a/(1+y^2)=0,-y+a/(1+x^2)=0 for x,y
\end{verbatim}
\end{widetext}
into Wolfram Alpha.
In this case, there are three real solutions. One of them is just that
given in EQ. \ref{solution1} and is real for all values of $a$.
The other two solutions are:
\begin{equation}
x=\frac{1}{2} \left ( a+\sqrt{a^2-4} \right ),~
y=\frac{1}{2} \left ( a-\sqrt{a^2-4} \right )
\label{no1}
\end{equation}
or
\begin{equation}
x=\frac{1}{2} \left ( a-\sqrt{a^2-4} \right ),~
y=\frac{1}{2} \left ( a+\sqrt{a^2-4} \right ),
\label{no2}
\end{equation}
both of which are real only for $a>2$.
These two solutions are the bistable solutions:  either $x$ is large
and $y$ is small (EQ. \ref{no1}) or $y$ is large and $x$ is small (EQ. \ref{no2}).
All three real solutions are plotted together in Fig. \ref{ROOTS} in magenta, red and
blue.
For the magenta solution, $x=y$. For the second solution,
$x$ is the blue line and  $y$ is the red line. For the third solution,
$x$ is the red line and $y$ is the blue line. 

\subsection{Stable or unstable?}
To determine whether or not a particular solution is stable, we consider
concentrations that are slightly different from the steady-state solutions
that we just found, {\em i.e.}, we set
\begin{equation}
x = x^* +\delta
\label{D1}
\end{equation}
and
\begin{equation}
y = y^* +\epsilon,
\label{D2}
\end{equation}
where $x^*$ and $y^*$ are the
steady state solutions that we just found and $\delta$ and $\epsilon$
are small deviations. For a stable solution, $\delta$ and $\epsilon$ will
evolve to smaller values in time. For an unstable solution,
$\delta$ and $\epsilon$ will
evolve to larger values in time.

To determine the time evolution of $\delta$ and $\epsilon$,
we substitute EQ. \ref{D1} and EQ. \ref{D2}
into EQ. \ref{RATE1B} and EQ. \ref{RATE2B} (with $n=2$) to obtain at linear order
in $\delta$ and $\epsilon$.
The results are:
\begin{equation}
\frac{1}{K} \frac{{\rm d\,} \delta}{{\rm d\,}t} = 
-\delta - Y \epsilon,
\label{RATE1d}
\end{equation}
where  
\begin{equation}
Y= \frac{2ay^*}{\left[1+(y^*)^2\right]^2},
\end{equation}
and, similarly
\begin{equation}
\frac{1}{K} \frac{{\rm d\,} \epsilon}{{\rm d\,}t} =  -\epsilon - X \delta,
\label{RATE2d}
\end{equation}
where
\begin{equation}
X= \frac{2ax^*}{\left[1+(x^*)^2\right]^2}.
\end{equation}

EQ. \ref{RATE1d} and EQ \ref{RATE2d} are similar to equations that
describe coupled harmonic
oscillators, that the students have seen earlier in the year.
Consequently, they are comfortable that these
equations realize normal modes, characterized by
eigenvalues and eigenvectors.
 Since we are interested in whether $\delta$ and $\epsilon$ shrink or grow versus
time, the key quantities that we need to find are the eigenvalues,
which report upon the decay (or growth) rates for
particular linear combinations of $\delta$ and $\epsilon$.
To determine the eigenvalues,
we assume that $\delta$ and $\epsilon$ decay
exponentially decay in time,
{\em i.e.}, we assume that $\delta = D e^{-\Gamma t}$ and
$\epsilon= E e^{-\Gamma t}$. We then
substitute these guesses into
EQ.  \ref{RATE1d} and EQ. \ref{RATE2d}, and
solve for $\Gamma$ in terms of the parameters of the problem.
Finally, we decide whether $\Gamma$ is positive, corresponding to $\delta$ and $\epsilon$
that decrease in time
and therefore a stable solution,
or whether $\Gamma$ is negative, corresponding to an unstable solution.

Following the first step of this procedure, we find
\begin{equation}
-\frac{\Gamma}{K} {\delta}=  -\delta - Y \epsilon.
\label{RATE2e}
\end{equation}
\begin{equation}
-\frac{\Gamma}{K} {\epsilon} =  -\epsilon - X \delta,
\end{equation}
whence
\begin{equation}
\Gamma = K(1 \pm \sqrt{Y X}).
\label{GAMMA}
\end{equation}
EQ. \ref{GAMMA} is applicable to all of the solutions we have found
previously. It is simply necessary to use the appropriate values of $Y$ and $X$.

First, we examine the bistable solutions. Using the expressions given
in EQ. \ref{no1} and EQ. \ref{no2}, we find
\begin{equation}
\Gamma = K\left(1\pm \frac{2}{a}\right).
\label{GAMMABISTABLE}
\end{equation}
Recalling that the bistable solutions are real only for $a>2$,
EQ. \ref{GAMMABISTABLE} informs us that the
values of $\Gamma$ corresponding to the two eigenmodes of the bistable
solutions are both invariably positive. Consequently, they both correspond to
decaying exponentials, and we see that the bistable solutions are, in fact, stable.

\begin{figure}[t!]
\begin{center}
\includegraphics[scale=1.00]{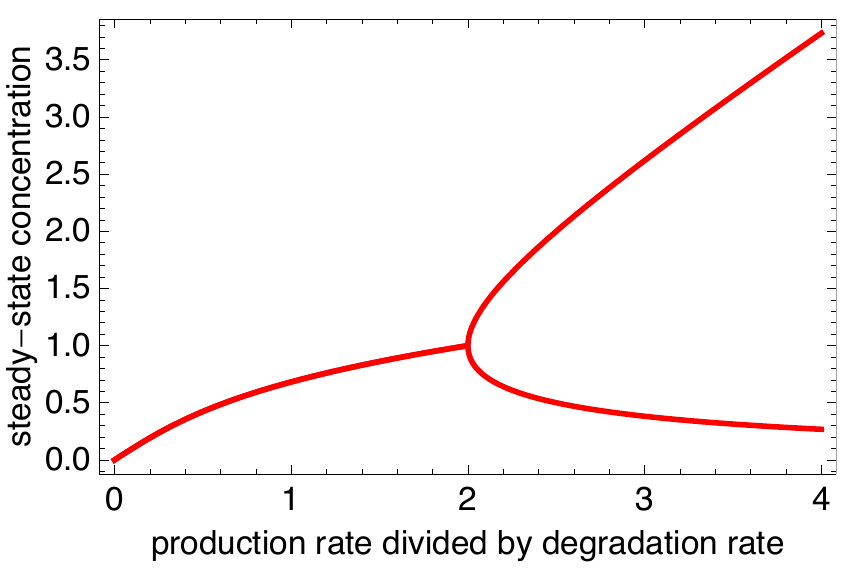}
\end{center}
\caption{
Steady-state concentrations for the genetic toggle switch
plotted versus the ratio of the production rate divided by the degradation rate.
}
\label{PHASEDIAGRAM}
\end{figure}

For the other real  solution, corresponding to the magenta curve in Fig. \ref{ROOTS}, we have
$x^*=y^*$, so that $X=Y$, and
\begin{equation}
\Gamma = K(1\pm Y).
\label{GAMMANONBISTABLE}
\end{equation}
It is straightforward to show that  $Y$ is less than unity for $a<2$ and
greater than unity for $a>2$. It follows that for $a<2$, $\Gamma$ is positive
corresponding to a stable solution. However, for $a>2$,
 $\Gamma$ is
negative. 
We only need one of the eigenvalues to be negative to send us away from the
steady-state solution. Therefore, this is indeed an unstable
solution, and it is not realized, because any small
fluctuation grows away from it.
Such a fluctuation away from the unstable solution will eventually approach one of
the stable solutions, as is apparent from the Mathematica demonstration.
Finally, then, we can report the ``equation of state'' of the genetic
toggle switch in Fig. \ref{PHASEDIAGRAM}. For $a<2$, there is   a single
solution with $c_1=c_2$, {\em i.e.}, there is no bistability.
By contrast, for $a>2$, there are two bistable
solutions with $x>y$ or $y > x$.
Importantly, whether or not bistable behavior is realized depends on the parameters of the model.
In class, we discuss that the model's prediction of bistable behavior
is just what is observed in the experiments of Ref.~\onlinecite{GTS}.

\section{The Repressilator}
\begin{figure}[h!]
\begin{center}
\includegraphics[width=0.35\textwidth]{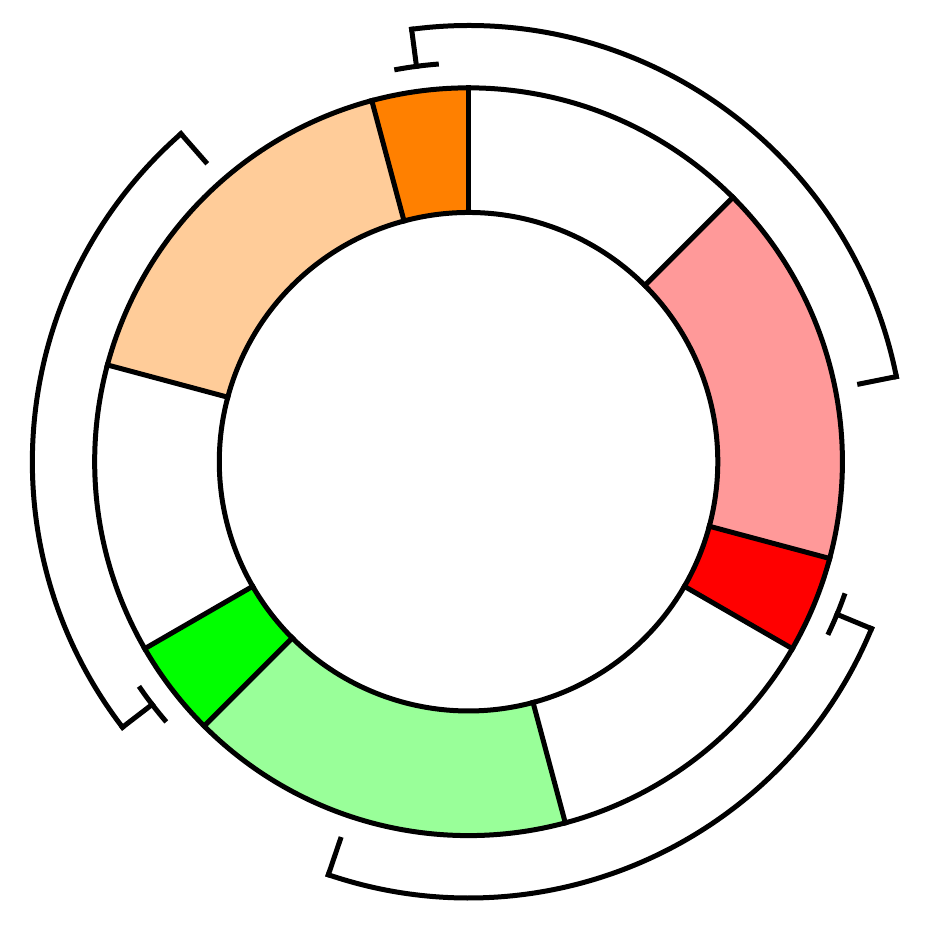}
\end{center}
\caption{ 
Schematic of the repressilator plasmid.
The solid orange region is the promoter for repressor 1.
The solid green region is the promoter for repressor 2.
The solid red region is the promoter for repressor 3.
The light orange region codes for repressor 1, which, as indicated, represses expression of repressor 2.
The light green region codes for repressor 2, which, as indicated, represses expression of repressor 3.
The light red region codes for repressor 3, which, as indicated, represses expression of repressor 1.
}
\label{REPRESSILATOR}
\end{figure}

Fig. \ref{REPRESSILATOR} shows the genetic architecture of the repressilator.
This gene circuit is composed of three genes and their gene products, each of which represses
expression of the following gene's gene product. This circuit uses the same basic element as the gene toggle switch, but uses three of them, rather
than two. The odd number of elements going around the complete circuit
gives rise to negative feedback.
We will see that negative feedback around a circuit can lead to
oscillations.

%

\begin{figure}
\begin{center}
\includegraphics[width=3in]{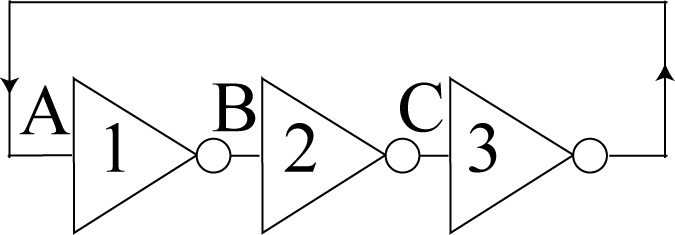}
\vskip 0.5 in
\includegraphics[width=2in]{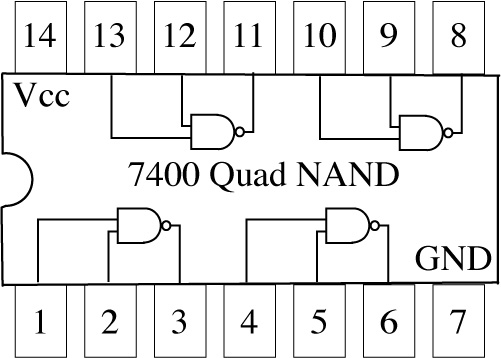}
\vskip 0.5 in
\includegraphics[width=3.5in]{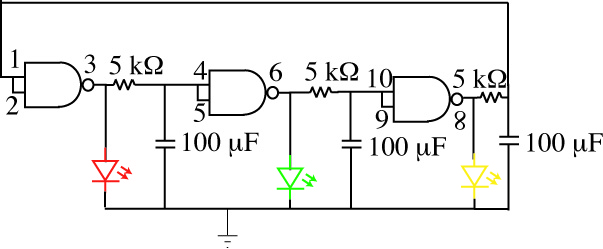}
\end{center}
\caption{Top: Three inverters connected so that the output of the third inverter is
routed to the input of the first inverter.
Middle: Quad NAND gate chip used to produce three inverters.
Bottom: Three inverters  with resistors and capacitors incorporated in the circuit
to achieve a convenient oscillation period.
}
\label{ThreeInverters}  
\end{figure}

\subsection{An electronic repressilator}
In the PHYS 165/166 laboratory class, students are invited to analyze the Boolean logic when
three inverters are connected as shown at the top of Fig. \ref{ThreeInverters}.
In this case,
if  input A is {\tt True}, then B is {\tt False}, then C is {\tt True}, then A is {\tt False}. 
But this configuration is inconsistent, because our original assumption was that  input A was {\tt True}. 
Do things work out better if we start off with input A {\tt False}?
Then B is {\tt True}, then C is {\tt False}, then A  is {\tt True}.
The logic seems to fail in both cases. 
The resolution of this contradiction is to admit that in any actual circuit, whether it is genetic, electronic,
or mathematical, it will take a non-zero period of time to switch from {\tt True} to {\tt False} and {\em vice versa}.
Then, it turns out, the ``logic'' succeeds:  Each inverter will switch states, then the next, then the next, then
the first again, {\em ad infinitum}.   In this way, we may realize an oscillator.

\begin{figure}[t!]
\begin{center}
\includegraphics[scale=0.07]{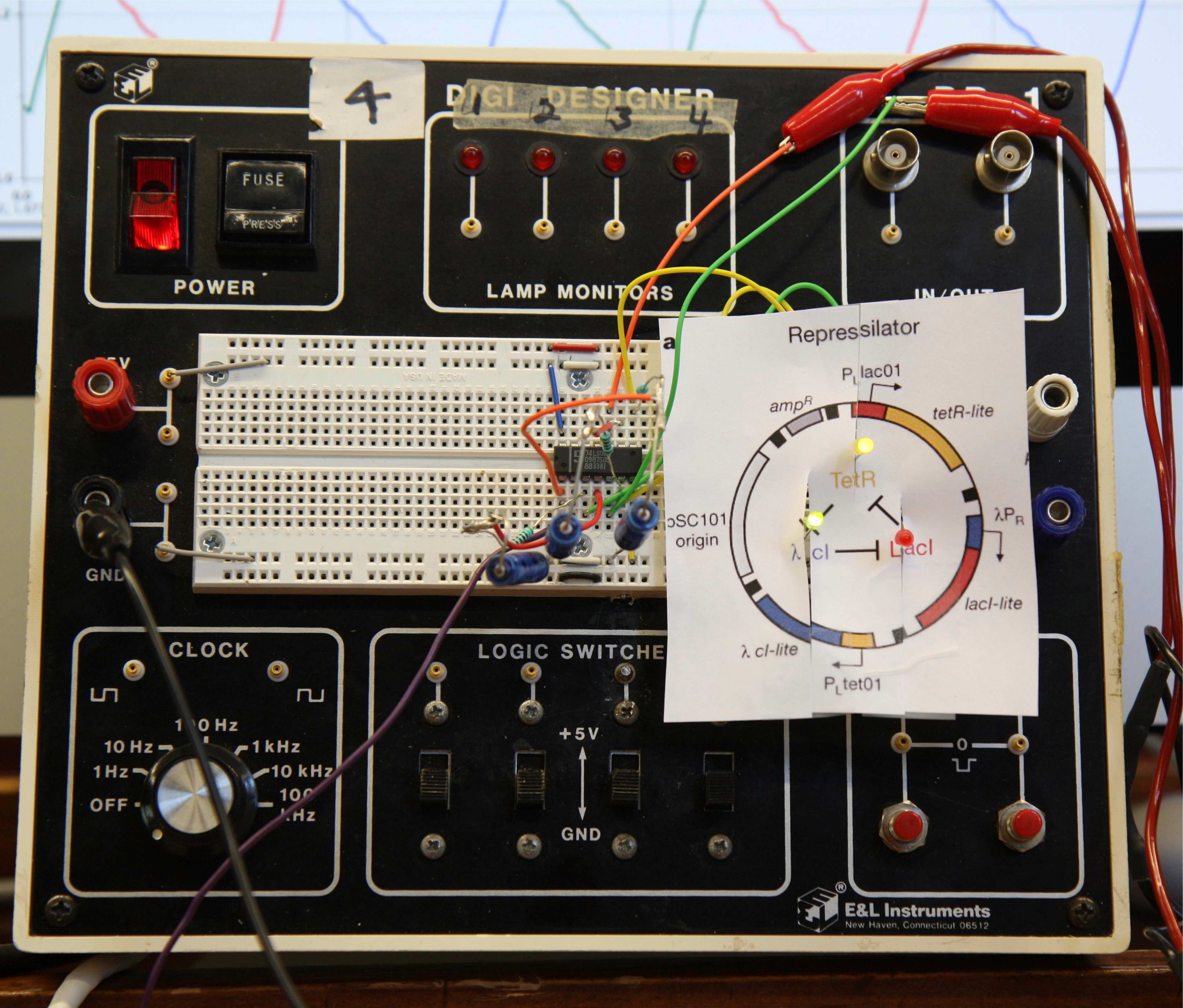}
\includegraphics[scale=0.14]{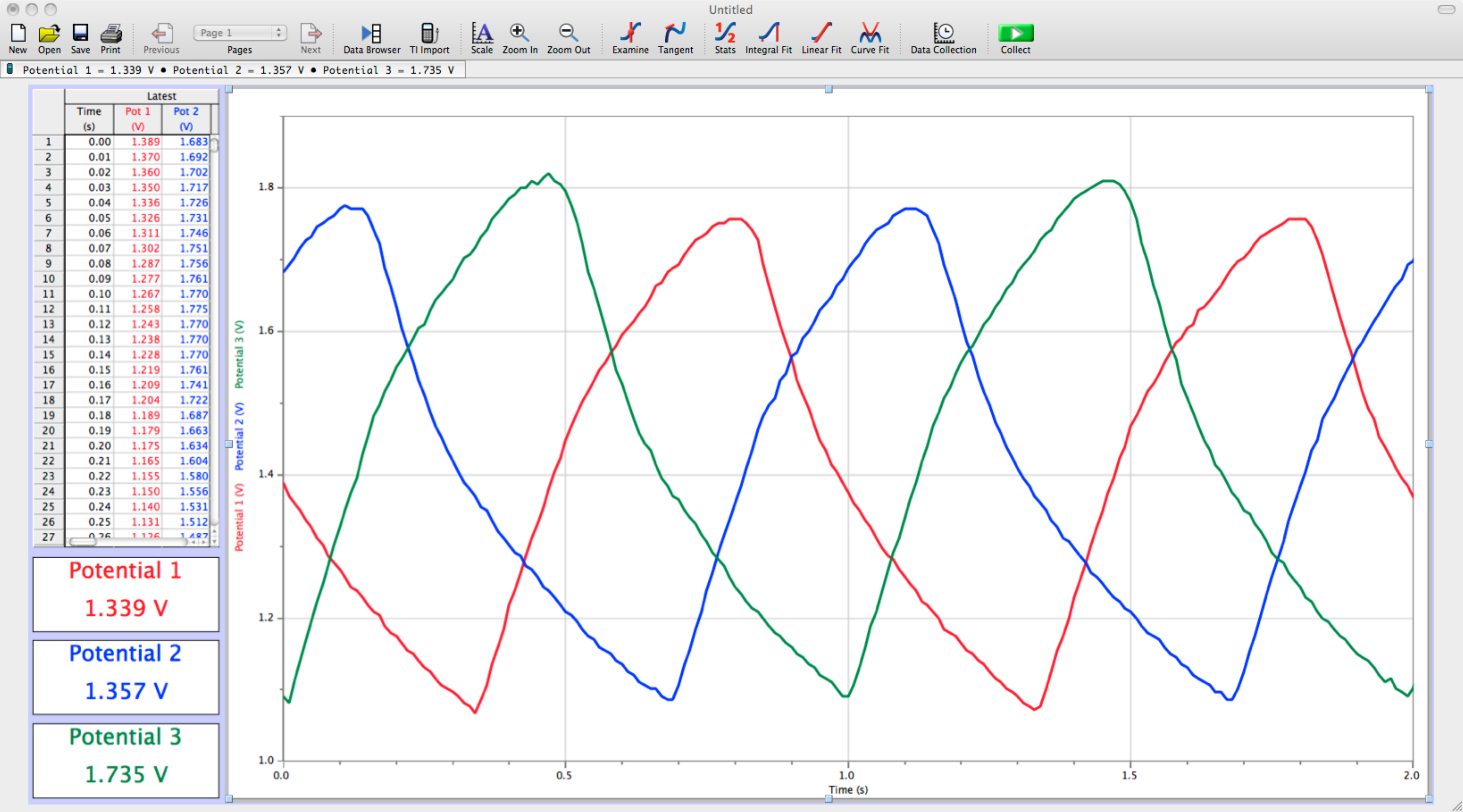}
\end{center}
\caption{
Photograph of our  implementation of an electronic repressilator (top)
together with the voltage outputs observed on an oscilloscope (bottom).
The shape or voltage outputs
is strikingly similar to the
Mathematica traces in Fig. \ref{RepressScreenShot}
}
\label{BOTH}
\end{figure}

We realize an electronic version of the  repressilator using three NAND gates of 
a 7400 quad NAND chip (Fig. \ref{ThreeInverters}).
To give the circuit  a convenient oscillation period,
we introduced capacitors and resistors as shown at the bottom of Fig. \ref{ThreeInverters}.
Because the capacitor takes a finite amount of time to charge or discharge through the resistor
after the output of each NAND gate has switched,
a delay is incorporated between the time at which the output of each NAND gate
switches and the time at which the input to the next NAND gate reaches the
threshold voltage for switching the next NAND gate.
The specific values used are $R=5$  k$\Omega$ and $C=100$  $\mu$F, which yield an
oscillation period of approximately 1~s.
To provide a vivid readout of the state of the repressilator circuit, red, yellow and green LEDs are introduced (see Fig. \ref{ThreeInverters}).
As shown in Supplementary Movie 1  on YouTube, \href{http://www.youtube.com/watch?v=S-Ktb5SFgv4}{\tt http://www.youtube.com/v/sKtb5SFgv4},
they oscillate happily with the advertised period.  
In the laboratory class, students are asked to
examine the signals on an oscilloscope attached to the inputs of the NAND chip, pins 1,4, and 9,
and determine
more precisely the period of an entire cycle of flashing lights,
as shown in Fig.~\ref{BOTH}, which also
illustrates the breadboard implementation of our electronic repressilator.  

\subsection{Chemical rate equations for the repressilator}
Our analysis of the repressilator builds directly on the previous analysis of the genetic toggle switch.
Consequently, the analysis presented here is simpler and, we believe, more accessible to our
students than that given in Ref.~\onlinecite{REPRESS}.
For the repressilator, we simply add a third gene, so that the relevant chemical rate equations
become:
\begin{equation}
\frac{1}{K} \frac{{\rm d\,}x_1}{{\rm d\,}t} =  - x_1 +  \frac{a}{1+x_2^n},
\label{RATE99A}
\end{equation}
\begin{equation}
\frac{1}{K} \frac{{\rm d\,}x_2}{{\rm d\,}t} =  -x_2 + \frac{a}{1+x_3^n} ,
\label{RATE99B}
\end{equation}
and
\begin{equation}
\frac{1}{K} \frac{{\rm d\,}x_3}{{\rm d\,}t} =  -x_3 + \frac{a}{1+x_1^n},
\label{RATE99C}
\end{equation}
where now $x_1=c_1/c_0$, {\em etc.}
In this case, we will leave $n$ as is.
The reason is that it will turn out that this system does not realize sustained
oscillations for $n \leq 2$ for any
value of $a$. We will determine an approximate ``phase diagram'' of the repressilator as
as function of $a$ and $n$.

\begin{figure}[b!]
\vspace{-0.1in}
\begin{center}
{\includegraphics[width=0.48\textwidth,keepaspectratio=true]{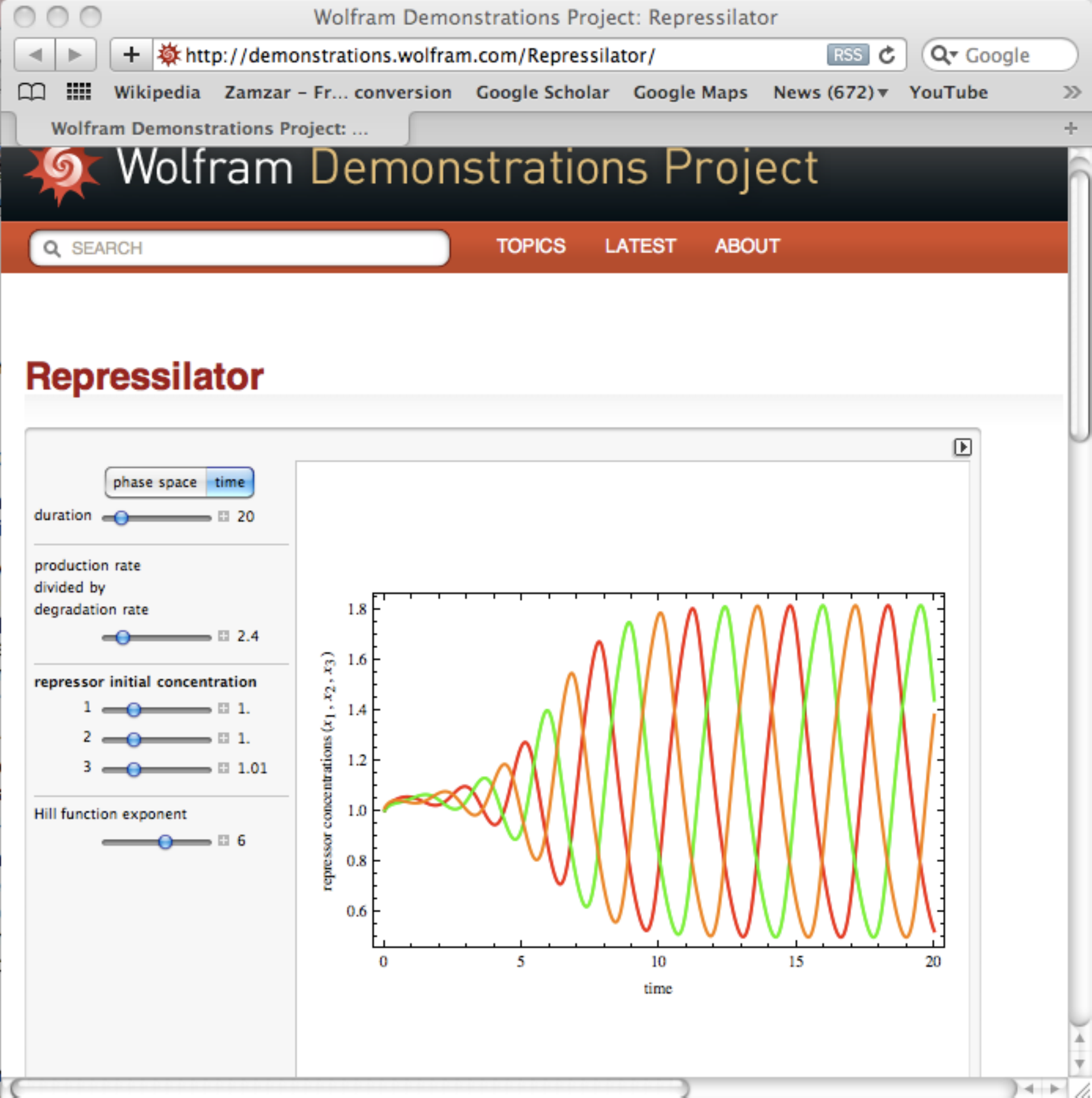}}
\end{center}
\vspace{-0.2in}
\caption{
Screenshot, showing our Mathematica Demonstration  \cite{repressdemo}
of the behavior of the repressilator  for $a=12$, $n=6$, $x_1(0)=2$,
$x_2(0)=2$, and $x_3(0)=4$.
To investigate the solutions on their own,
students are able to adjust the parameters of the simulation, using the sliders and toggles,
directly from their web browsers.
}
\label{RepressScreenShot}
\vspace{-0.1in}
\end{figure}

\subsection{Mathematica solutions for the repressilator}
This system of equations (EQ. \ref{RATE99A}, EQ. \ref{RATE99B}, and EQ. \ref{RATE99C}) can be solved numerically by Mathematica, and 
we have written and published a
\href{http://demonstrations.wolfram.com/Repressilator/}{Wolfram Demonstration} that does just this.\cite{repressdemo}
A screen shot showing $x_1$, $x_2$, and $x_3$ as a function of time is
presented in Fig. \ref{RepressScreenShot}.
 For certain parameter values, the numerical solution approaches
a steady-state value for the concentrations after decaying oscillatory behavior at early times.
For other parameter values, including those corresponding to Fig. \ref{RepressScreenShot},
the numerical
solution shows sustained, constant-amplitude oscillations. 
In this latter case, this is a biologic clock, which might be a model for the clock that operates
in cell division, or establishes a circadian rhythm. 
These sustained oscillations are not a steady-state solution, but they are the interesting
solution in this case. Therefore, in this case, we want to find  the range of parameter
values for which such unstable, oscillatory solutions are observed. 
Nevertheless, our procedure in this case will mirror the procedure
that we followed in the case of the genetic toggle switch:
We  will
find the steady-state solutions.
Next, we will derive the equations that describe how small deviations ($\delta_1$, $\delta_2$, and $\delta_3$)
in concentration from the steady state
values evolve in time.
Then, we will assume an exponential time dependence for $\delta_1$, $\delta_2$, and $\delta_3$
and solve for the corresponding values of  $\Gamma$ in terms of the parameters of the problem.
Where $\Gamma$ is positive corresponds to a stable solution.
Where $\Gamma$ is negative corresponds to an unstable solution, which in this case, 
is the more interesting solution.

\subsection{Steady-state solutions for the repressilator}
In a steady-state, EQ.  \ref{RATE99A}, EQ. \ref{RATE99B} and EQ. \ref{RATE99C} become
\begin{equation}
x^*_1 =  \frac{a}{1+(x^*_2)^n},
\label{RATE99D}
\end{equation}
\begin{equation}
 x^*_2 = \frac{a}{1+(x^*_3)^n} ,
\label{RATE99E}
\end{equation}
and
\begin{equation}
 x^*_3 = \frac{a}{1+(x^*_1)^n},
\label{RATE99F}
\end{equation}
where the $^*$ indicates the steady-state value.
These equations initially seem daunting to solve, but exploration of the numerical solution,
given in the Mathematica demonstration, does not reveal any steady-state solutions
for which the concentrations ($x_1$, $x_2$, and $x_3$) are different from each other. 
Therefore, in
fact, all of the real solutions to EQ.  \ref{RATE99A}, EQ. \ref{RATE99B} and EQ. \ref{RATE99C} correspond to $x^*_1=x^*_2=x^*_3= x^*$, say.
(Since concentrations must be real, we are interested solely in real solutions.)
In this case,  each of EQ. \ref{RATE99D}, EQ. \ref{RATE99E}, and EQ. \ref{RATE99F} reduces to
\begin{equation}
   x^* =  \frac{a}{1+(x^*)^n}.
\label{RATE100}
\end{equation}
It is straightforward to see graphically that EQ. \ref{RATE100} has only one real solution:
we plot $y = x^*$ and $y={a}/[{1+(x^*)^n]}$, and  where these
two curves cross corresponds to the real solutions for $x^*$ that we seek.
It is clear from this plot (not shown)
that there is always one and only one intersection point,  and therefore one and only one
real solution,
 irrespective of the value of $a$.


\subsection{Stability analysis for the repressilator}
To examine the stability of this solution, we proceed similarly to above.
That is, we assume
$x_1 = x^*+\delta_1$,
$x_2=x^*+\delta_2$,
$x_3=x^*+\delta_3$, and
carry out a  linear expansion of
EQ. \ref{RATE99A}, \ref{RATE99B}, and \ref{RATE99C}
leading to
\begin{equation}
\frac{1}{K} \frac{{\rm d\,} \delta_1}{{\rm d\,}t} =  -\delta_1 - X_n \delta_2,
\label{RATE3d}
\end{equation}
\begin{equation}
\frac{1}{K} \frac{{\rm d\,} \delta_2}{{\rm d\,}t} =  -\delta_2 - X_n \delta_3,
\label{RATE3e}
\end{equation}
and
\begin{equation}
\frac{1}{K} \frac{{\rm d\,} \delta_3}{{\rm d\,}t} =  -\delta_3 - X_n \delta_1,
\label{RATE3f}
\end{equation}
where
\begin{equation}
X_n = \frac{n\left(x^*\right)^{n-1}a}{\left(1+\left(x^*\right)^n\right)^2}.
\label{YYY}
\end{equation}
Next,
we assume that $\delta_1 = D_1 e^{-\Gamma t}$, 
$\delta_2 = D_2 e^{-\Gamma t}$, and
$\delta_3 = D_3 e^{-\Gamma t}$,
with the result that
\begin{equation}
{\Gamma} =K\left[1+X_n\right],
\label{Y4a}
\end{equation}
or
\begin{equation}
{\Gamma} =K\left[1+\frac{X_n}{2} + i\frac{\sqrt{3}X_n}{2}\right],
\label{Y4b}
\end{equation}
or
\begin{equation}
{\Gamma} =K\left[1-\frac{X_n}{2} + i\frac{\sqrt{3}X_n}{2}\right].
\label{Y4c}
\end{equation}
Since $X_n$ is  positive, EQ. \ref{Y4a} corresponds to an exponential decay.
Since $1+X_n/2$ is  also  positive, EQ. \ref{Y4b} 
corresponds to an exponentially decaying oscillation. Therefore, 
both of these eigenvalues  correspond to stable solutions.
EQ. \ref{Y4c} is stable for $X_n<2$. However,  for $X_n > 2$,  $1 - X_n/2$ is negative,
and this solution
grows exponentially in an oscillating fashion. Therefore,
the condition to realize an unstable solution  is $X_n >2$,
which in fact corresponds to sustained oscillations.

\subsection{``Phase diagram'' for the repressilator}
\begin{figure}[t!]
\begin{center}
\includegraphics[width=3.2in]{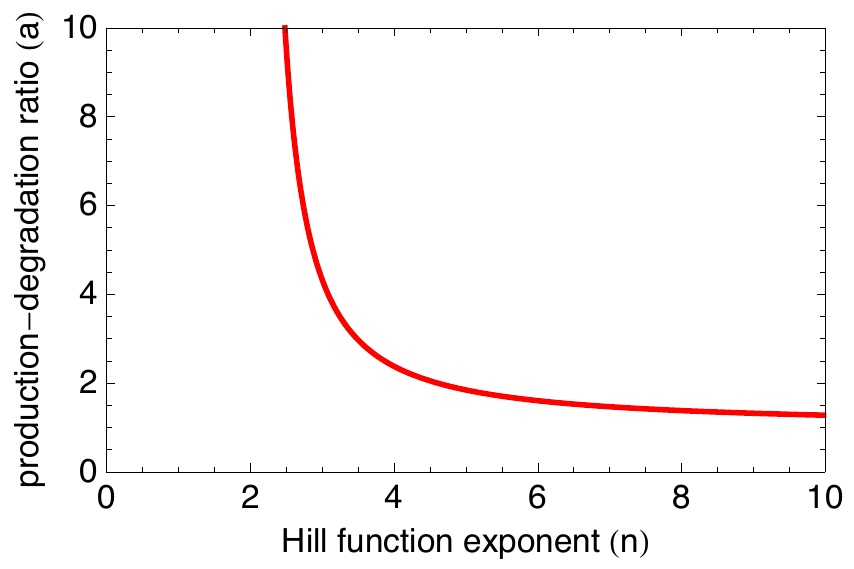}
\end{center}
\caption{
Approximate repressilator phase diagram.
The region above the red line corresponds to sustained oscillations -- a limit cycle.
The region below the red line corresponds to stable solutions for which
$x_1^*=x_2^*=x_3^*$.
}
\label{RPD}
\end{figure}
What is the value of $X_n$? So far, we have
not calculated an explicit value for $X_n$, because we have
not calculated an explicit value for $x^*$.
This is because it is not possible to write down the solution of EQ. \ref{RATE100}
analytically for arbitrary $n$.  However,
we can find a useful, approximate solution as follows.
First we note that using EQ.~\ref{RATE100},
EQ.~\ref{YYY} may be rewritten:
\begin{equation}
X_n = \frac{n\left(x^*\right)^{n+1}}{a}.
\label{YYY2000}
\end{equation}
We may also re-write EQ.~\ref{RATE100} as
\begin{equation}
(x^*)^{n+1} + x^* = a.
\label{RATE10000}
\end{equation}
If we assume that $(x^*)^n \gg 1$,
it follows from EQ.~\ref{RATE10000} that
\begin{equation}
(x^*)^{n+1} \simeq a-a ^{\frac{1}{n+1}}.
\label{ABAB}
\end{equation}
Combining EQ. \ref{YYY2000} and EQ. \ref{ABAB}, we find
\begin{equation}
X_n \simeq n(1-a^{-n/(n+1)}),
\label{YOY}
\end{equation}
which represents a useful approximate expression for $X_n$ that we can
use to determine the stability of the steady state solution.
First, notice that for $n=2$, $X_2$ is inevitably less than 2.
This result survives an exact calculation, which is possible for $n=2$.
Therefore, there are no sustained oscillations for $n=2$.
It's a different story for $n>2$.
In this case, according to EQ. \ref{Y4c} the steady state solution is unstable for 
\begin{equation}
n(1-a^{-n/(n+1)}) > 2.
\end{equation}
After some algebra, this condition becomes that the steady state solution is unstable for
\begin{equation}
a>\left(1-\frac{2}{n}\right)^{-\left(1+\frac{1}{n}\right)}.
\label{REPRESS_PHASE_DIA}
\end{equation}
EQ. \ref{REPRESS_PHASE_DIA} represents a ``phase diagram'' for the repressilator, 
specifying the region of $n$-$a$ ``phase space''
in which the steady-state solution is unstable and which therefore realizes sustained oscillations. This (approximate) repressilator
phase diagram is show in Fig. \ref{RPD}.
The region above the  line corresponds to sustained oscillations -- a limit cycle.
The region below the line corresponds to a stable fixed point. Fig. \ref{RPD} is approximate because EQ. \ref{YOY} is approximate.
Nevertheless, when the students run the repressilator Mathematica demonstration, they
are soon able to convince themselves that our analytic phase diagram is 
qualitatively correct. Specifically, they find that  to realize sustained oscillations requires 
larger values of $a$ at smaller values of $n$
and relatively smaller values of $a$ at larger values of $n$.
Again, we see that the behavior realized depends on
biochemical parameters of the model.
And again, in class, we discuss that the
model's prediction of oscillatory behavior
is what is observed in the experiments of
Ref.~\onlinecite{REPRESS}.

\section{Summary}
We have discussed and analyzed two prototypical gene circuits with feedback,
namely  the genetic toggle switch and the repressilator, which together constitute the final topic in
a year-long introductory physics sequence for biology and pre-medical students at Yale.
Our analytic, numerical, and electronic treatments of the genetic toggle switch,
which consists of two genes, whose protein products each represses the other's gene expression, 
reveals that this circuit realizes bistability.
Our new, simplified treatment of the repressilator,
which consists of three genes, each of whose protein product represses the next gene's expression,
reveals that this circuit realizes
sustained oscillations for certain parameter values.
In both cases, we obtained a ``phase diagram'' that specifies the region of parameter space
in which bistability or oscillatory behavior, respectively, are realized.

\acknowledgments
We thank the PHYS 170/171 and PHYS 165/166 classes for their participation,
and Sean Barrett, Ross Boltyanskiy, Diego Caballero, Rick Casten, Betsy Cowell,
Jane Cummings, Stefan Elrington, Merideth Frey,
Eric Holland, Syed Hussaini, Sohrab Ismail-Beigi,
Anna Kashkanova, Peter Koo, Andrew Mack, Wambui Muturi, Rona Ramos,
Raphael Sarfati, William Segraves, Gennady Voronov,
Christine Willinger, and Yao Zhao for valuable discussions.
SGJM acknowledges support from the NSF via PHY 1019147.
\bibliographystyle{./aip}
\bibliography{./simon}

\begin{thebibliography}{10}

\bibitem{BIO2010}
National~Research Council,
\newblock {\em BIO2010: Transforming Undergraduate Education for Future
  Research Biologists},
\newblock National Academies Press, Washington, DC, 2003.

\bibitem{AAMC}
Association of~American Medical Colleges~(AAMC) and the Howard Hughes Medical
  Institute~(HHMI),
\newblock {\em Scientific Foundations for Future Physicians},
\newblock 2009.

\bibitem{Handelsman2007}
Jo~Handelsman, Sarah Miller, and Christine Pfund,
\newblock {\em Scientific Teaching},
\newblock Freeman, New York, 2007.

\bibitem{Mochrie2011}
S.~G.~J. Mochrie,
\newblock ``The Boltzmann factor, {DNA} melting, and {Brownian} ratchets:
  Topics in an introductory physics sequence for biology and premedical
  students'',
\newblock American Journal of Physics {\bf 79}, 1121--1130 (2011).

\bibitem{GTS}
T.~S. Gardner C.~R. Cantor and J.~J. Collins,
\newblock ``Construction of a Genetic Toggle Switch in {\em Escherichia
  coli}'',
\newblock Nature {\bf 403}, 339--342 (2000).

\bibitem{REPRESS}
M.~B. Elowitz and S.~Leibler,
\newblock ``A Synthetic Oscillatory Network of Transcriptional Regulators'',
\newblock Nature {\bf 403}, 335--338 (2000).

\bibitem{Ptashne}
M.~Ptashne,
\newblock {\em A genetic switch: Phage Lambda revisited},
\newblock Cold Spring Harbor Press, Cold Spring Harbor, New York, 2004.

\bibitem{HolleyReview}
A.~Mara and S.~A. Holley,
\newblock ``Oscillators and the emergence of tissue organization during
  zebrafish somitogenesis'',
\newblock Trends in Cell Biology {\bf 17}, 593--599 (2007).

\bibitem{Schultz2009}
D.~Schultz, P.~G. Wolynes, E.~Ben Jacob, and J.~N. Onuchic,
\newblock ``Deciding fate in adverse times: Sporulation and competence in {\em
  Bacillus subtilis}'',
\newblock Proc. Nat. Acad. Sci. {\bf 106}, 21027--21034 (2009).

\bibitem{Dori2003}
Yehudit~Judy Dori, John Belcher, Mark Bessette, Michael Danziger, Andrew
  McKinney, and Erin Hult,
\newblock ``Technology for active learning'',
\newblock Materials Today {\bf 6}, 44--49 (2003).

\bibitem{Chabay2008}
R.~Chabay and B.~Sherwood,
\newblock ``Computational physics in the introductory calculus-based course'',
\newblock Am. J. Phys. {\bf 76}, 307--313 (2008).

\bibitem{DEMOSITE}
\htmladdnormallink{\tt
  <http://demonstrations.wolfram.com/>}{http://demonstrations.wolfram.com}.

\bibitem{gtsdemo}
\htmladdnormallink{\tt
  <http://demonstrations.wolfram.com/GeneticToggleSwitch/>}{http://demonstrations.wolfram.com/GeneticToggleSwitch/}.

\bibitem{repressdemo}
\htmladdnormallink{\tt
  <http://demonstrations.wolfram.com/Repressilator/>}{http://demonstrations.wolfram.com/Repressilator/}.

\bibitem{Pierre2008}
F.~St-Pierre and D.~Endy,
\newblock ``Determination of cell fate selection during phage lambda
  infection'',
\newblock Proc. Natl. Acad. Sci. USA {\bf 105}, 20705--10 (2008).

\bibitem{PBC}
R.~Philips amd J.~Kondev and J.~Theriot,
\newblock {\em Physical Biology of the Cell},
\newblock Garland, 2009.

\bibitem{WolframAlpha}
\htmladdnormallink{http://www.wolframalpha.com/}{http://www.wolframalpha.com/}.

\end{thebibliography}
\pagebreak

\end{document}